\pdfoutput=1
\documentclass[11pt]{article}

\usepackage[]{acl}

\usepackage{times}
\usepackage{latexsym}
\usepackage{adjustbox}
\usepackage{tikzducks}
\usepackage{graphicx}

\newcommand{\fon}[1]{\fontfamily{#1}\selectfont} 

\usepackage[T1]{fontenc}
\usepackage[utf8]{inputenc}

\usepackage{microtype}

\usepackage{inconsolata}
\usepackage{graphicx}
\usepackage{booktabs}
\usepackage{times}
\usepackage{latexsym}
\usepackage{svg}
\usepackage{amsmath}
\usepackage{tcolorbox}\tcbuselibrary{skins}
\usepackage{subcaption}
\usepackage{multirow}
\usepackage{array}
\usepackage{makecell}
\usepackage{float}
\usepackage{booktabs}
\usepackage{times}
\usepackage{latexsym}
\usepackage{graphicx}
\usepackage{url}
\usepackage{xcolor,colortbl}
\usepackage{soul}
\usepackage{color}
\usepackage{CJKutf8}
\usepackage{amsfonts}

\usepackage{array}
\newcolumntype{P}[1]{>{\centering\arraybackslash}p{#1}}

\usepackage{inconsolata}
\usepackage{graphicx}
\usepackage{booktabs}
\usepackage{times}
\usepackage{latexsym}
\usepackage{svg}
\usepackage{amsmath}
\usepackage{tcolorbox}\tcbuselibrary{skins}
\usepackage{subcaption}
\usepackage{multirow}
\usepackage{array}
\usepackage{makecell}
\usepackage{float}
\usepackage{booktabs}
\usepackage{times}
\usepackage{latexsym}
\usepackage{graphicx}
\usepackage{url}
\usepackage{xcolor,colortbl}
\usepackage{soul}
\usepackage{color}
\usepackage{CJKutf8}

\usepackage{enumitem}

\usepackage{hyperref}
\usepackage{url}
\usepackage{soul}
\usepackage{makecell}
\usepackage{threeparttable,subcaption}
\usepackage{tikz}
\usepackage{multirow}
\usepackage{booktabs,wrapfig}
\usepackage{amsthm}
\usepackage{setspace}
\usepackage{tcolorbox}\tcbuselibrary{skins}

\newtcolorbox{mybasecolorbox}[1][]{%
  colback=gray!25, colframe=gray!25,
  coltitle=black, fonttitle=\bfseries,
  sharp corners,
  width=(\linewidth-30pt),
  title=#1}

\def\tasknameLowercaseSpecial{\textit{backtracing}}
\def\tasknameLowercase{backtracing}
\def\tasknameUppercase{Backtracing}
\def\causalAcronym{ATE}
\def\gptLong{\texttt{\texttt{gpt-3.5-turbo-16k}}}
\def\lecture{\textsc{Lecture}}
\def\conv{\textsc{Conversation}}
\def\news{\textsc{News Article}}

\definecolor{CB_pear}{HTML}{BBCC33}
\definecolor{CB_pink}{HTML}{FFAABB}
\definecolor{CB_lightCyan}{HTML}{99DDFF}
\definecolor{CB_gray}{HTML}{DDDDDD}
\definecolor{CB_orange}{HTML}{EE8866}
\definecolor{CB_yellow}{HTML}{EEDD88}
\definecolor{text_green}{HTML}{CCDDAA}

\tcbset{sightsetting/.style={
    enhanced,
    size=fbox,
    boxrule=1.5pt,
    arc=2mm,
    auto outer arc,
    left=1pt,
    right=1pt,
    top=1pt,
    bottom=1pt,
    fontupper=\fon{lmss}, 
    colback=CB_lightCyan!15,
    colframe=CB_lightCyan!30,
    coltitle=CB_lightCyan!25!black, 
}}

\tcbset{dailydialogsetting/.style={
    enhanced,
    size=fbox,
    boxrule=1.5pt,
    arc=2mm,
    auto outer arc,
    left=1pt,
    right=1pt,
    top=1pt,
    bottom=1pt,
    colback=CB_pear!15,
    colframe=CB_pear!30,
    coltitle=CB_pear!25!black, 
    fontupper=\fon{lmss}, 
}}

\tcbset{inquisitivesetting/.style={
    enhanced,
    size=fbox,
    boxrule=1.5pt,
    arc=2mm,
    auto outer arc,
    left=1pt,
    right=1pt,
    top=1pt,
    bottom=1pt,
    fontupper=\fon{lmss}, 
    colback=CB_orange!10,
    colframe=CB_orange!25,
    coltitle=CB_orange!20!black, 
}}

\title{Backtracing: Retrieving the Cause of the Query}

\author{\vspace{.5em}{\bf Rose E. Wang}\quad {\bf Pawan Wirawarn}\quad {\bf Omar Khattab}\quad \\ \vspace{.5em} {\bf Noah Goodman}\quad {\bf Dorottya Demszky}\\ \vspace{.5em} Stanford University\\ \texttt{rewang@cs.stanford.edu, ddemszky@stanford.edu}}

\begin{document}
\maketitle

\begin{abstract}
Many online content portals allow users to ask questions to supplement their understanding (e.g., of lectures). 
While information retrieval (IR) systems may provide answers for such user queries, they do not directly assist content creators---such as lecturers who want to improve their content---identify segments that \textit{caused} a user to ask those questions.
We introduce the task of \emph{\tasknameLowercase{}}, in which systems retrieve the text segment that most likely caused a user query.
We formalize three real-world domains for which \tasknameLowercase{} is important in improving content delivery and communication: understanding the cause of (a) student confusion in the \textsc{Lecture} domain, (b) reader curiosity  in the \textsc{News Article} domain, and (c) user emotion in the \textsc{Conversation} domain.
We evaluate the zero-shot performance of popular information retrieval methods and language modeling methods, including bi-encoder, re-ranking and likelihood-based methods and ChatGPT.
While traditional IR systems retrieve semantically relevant information (e.g., details on ``projection matrices'' for a query ``does projecting multiple times still lead to the same point?''), they often miss the causally relevant context (e.g., the lecturer states ``projecting twice gets me the same answer as one projection''). 
Our results show that there is room for improvement on \tasknameLowercase{} and it requires new retrieval approaches.
We hope our benchmark serves to improve future retrieval systems for \tasknameLowercase{}, spawning systems that refine content generation and identify linguistic triggers influencing user queries.\footnote{
Our code and data are opensourced: \url{https://github.com/rosewang2008/backtracing}.}
\end{abstract}

\section{Introduction}

Content creators and communicators, such as lecturers, greatly value feedback on their content to address confusion and enhance its quality \citep{evans1978clarity, hativa1998lack}.
For example, when a student is confused by a lecture content, they post questions on the course forum seeking clarification. 
Lecturers want to determine \textit{where} in the lecture the misunderstanding stems from in order to improve their teaching materials \citep{mckone1999analysis, harvey2003student,gormally2014feedback}.
The needs of these \textit{content creators} are different than the needs of \textit{information seekers} like students, who may directly rely on information retrieval (IR) systems such as Q\&A methods to satisfy their information needs \citep{schutze2008introduction,yang2015wikiqa,rajpurkar2016squad,joshi2017triviaqa,yang2018hotpotqa}. 

\begin{figure}[t]
    \centering
    \includegraphics[width=\linewidth]{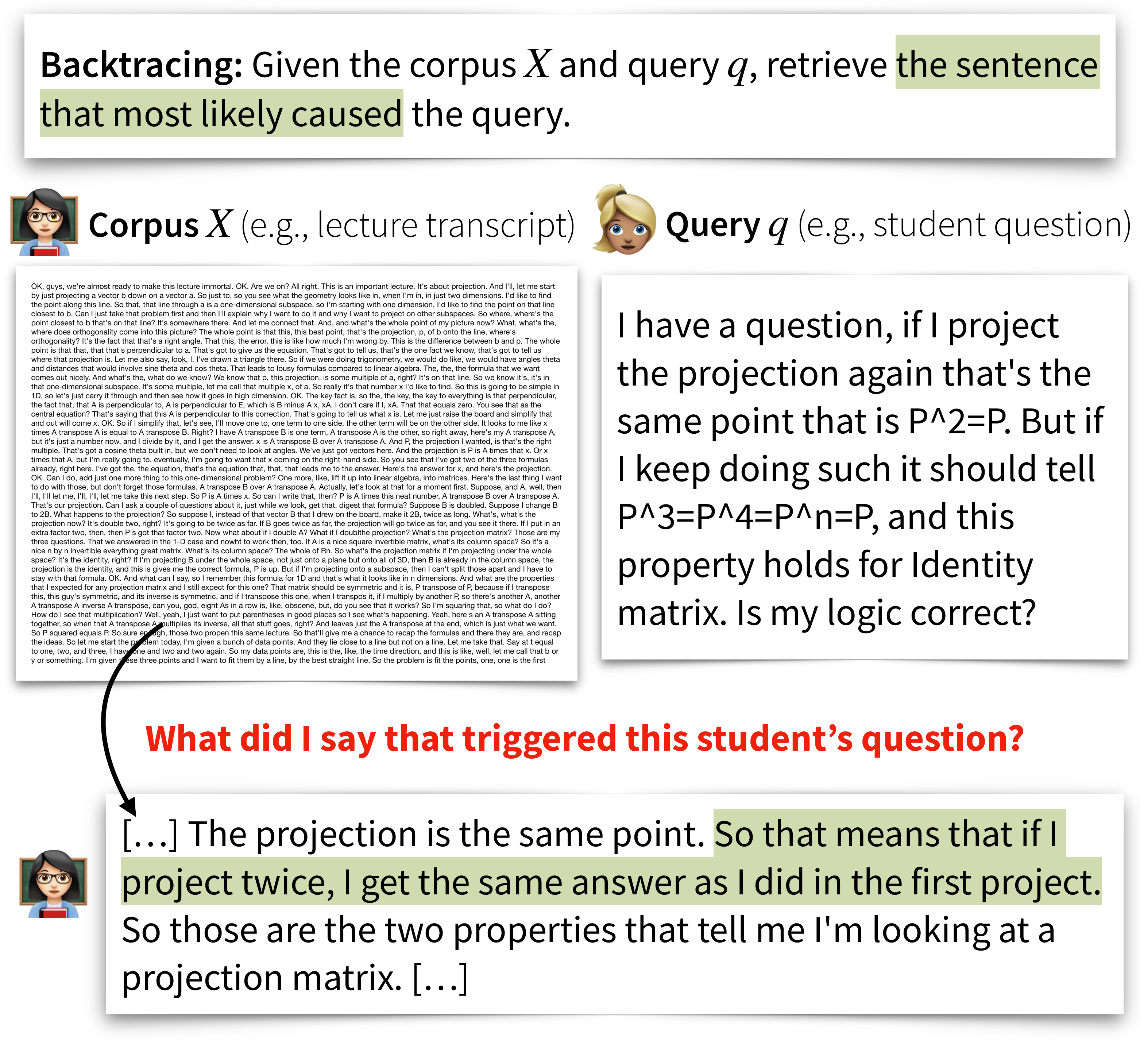}
    \vspace{-2em}
    \caption{
    \small
    The task of \tasknameLowercase{} takes a query and identifies the context that triggers this query. 
    Identifying the cause of a query can be challenging because of the lack of explicit labeling, large corpus size, and domain expertise to understand both the query and corpus. 
    % The task itself seems highly interesting. In principle, it is the inversion of deriving an information need from context: given an information need, identify the context that triggered this need (with a sentence as the level of detail). To me, the most valuable application for the task seem information retrieval systems, where a user's information need is expressed as a query (same as in the task presented here), which is usually underspecified. Knowing the reason that triggered the information need could provide additional details that help to provide relevant results to satisfy the information need.
    }
    \label{fig:fig1}
\end{figure}

\begin{figure*}[ht]
    \centering 
   \includegraphics[width=\linewidth]{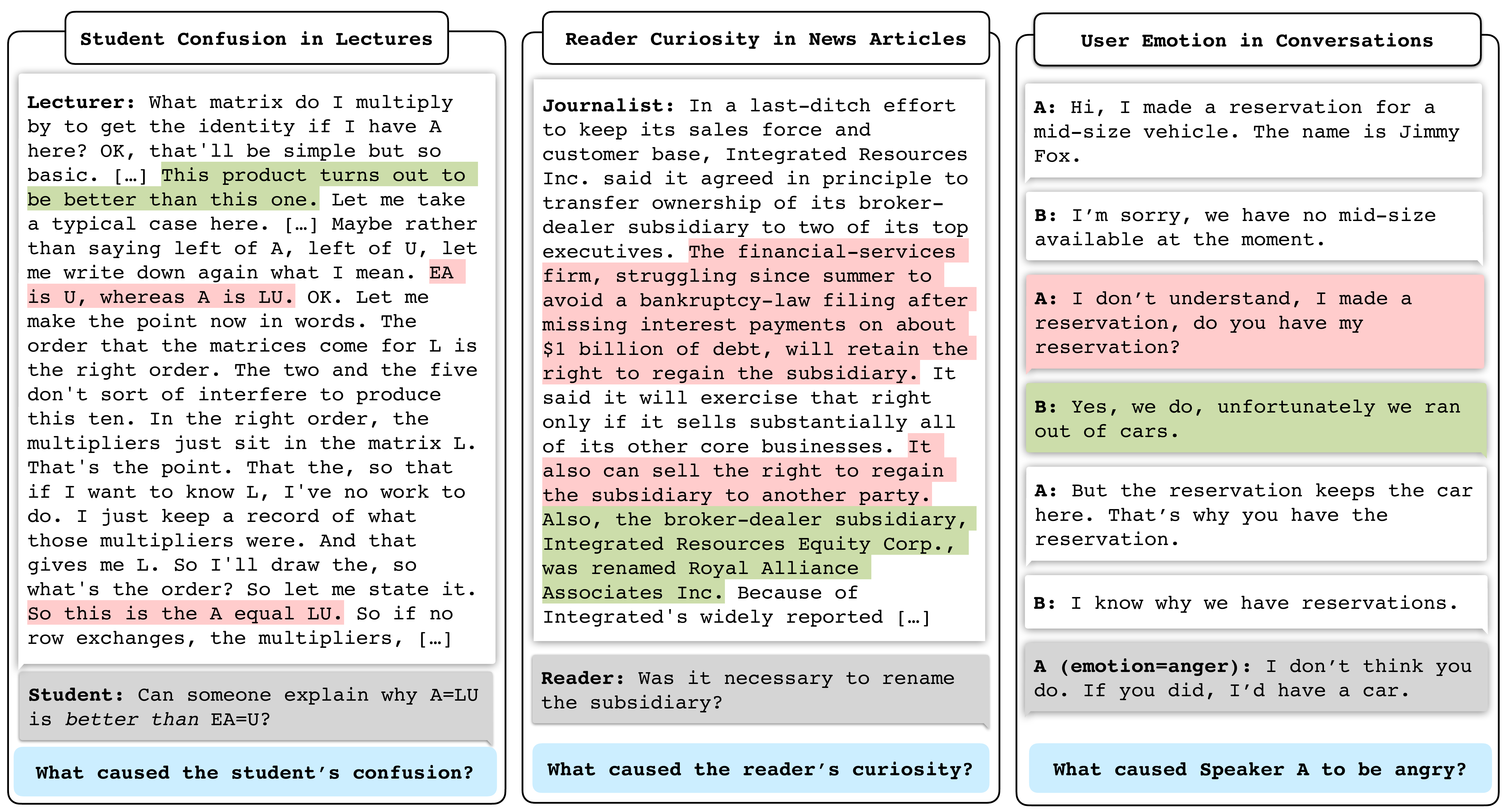}
   % \vspace{-2em}
    \caption{
    \label{fig:backassociation_examples}
    \small
    Retrieving the correct triggering context can provide insight into how to better satisfy the user's needs and improve content delivery.
    We formalize three real-world domains for which \tasknameLowercase{} is important in providing context on a user's query:
    (a) The \textsc{Lecture} domain where the objective is to retrieve the cause of student confusion;
    (b) The \textsc{News Article} domain where the objective is to retrieve the cause of reader curiosity; 
    (c) The \textsc{Conversation} domain where the objective is to retrieve the cause of user emotion (e.g., anger).
    The user's query is shown in the \colorbox{CB_gray}{gray} box and the triggering context is the \colorbox{text_green}{green}-highlighted sentence.
    Popular retrieval systems such as dense retriever-based and re-ranker based systems retrieve incorrect contexts shown in \colorbox{CB_pink}{red}.
    } 
\end{figure*}

Identifying the cause of a query can be challenging because of the lack of explicit labeling, implicit nature of additional information need, large size of corpus, and required domain expertise to understand both the query and corpus. 
Consider the example shown in Figure~\ref{fig:fig1}.
First, the student does not explicitly flag what part of the lecture causes their question, yet they express a latent need for additional information outside of the lecture content.
Second, texts like lecture transcripts are long documents; a lecturer would have a difficult time pinpointing the precise source of confusion for every student question they receive.
Finally, some queries require domain expertise for understanding the topic and reason behind the student's confusion; not every student question reflects the lecture content verbatim, which is what makes \tasknameLowercase{} interesting and challenging.

To formalize this task, we introduce a novel retrieval task called \tasknameLowercaseSpecial. Given a query (e.g., a student question) and a corpus (e.g., a lecture transcript), the system must identify the sentence that most likely provoked the query.
We formalize three real-world domains for which \tasknameLowercase{} is important for improving content delivery and communication.
First is the \textsc{Lecture} domain where the goal is to retrieve the cause of student confusion; the query is a student's question and the corpus is the lecturer's transcript. 
Second is the \textsc{News Article} domain where the goal is to retrieve the cause of a user's curiosity in the news article domain; the query is a user's question and the corpus is the news article. 
Third is the  \textsc{Conversation} domain where the goal is to retrieve the cause of a user's emotion (e.g., anger); the query is the user's conversation turn expressing that emotion and the corpus is the complete conversation. 
Figure~\ref{fig:backassociation_examples} illustrates an example for each of these domains.
These diverse domains showcase the applicability and common challenges of \tasknameLowercase{} for improving content generation, similar to heterogeneous IR datasets like BEIR \citep{thakur2021beir}. 

We evaluate a suite of popular retrieval systems, like dense retriever-based \citep{reimers2019sentence,guo2020multireqa,karpukhin-etal-2020-dense} or re-ranker-based systems \citep{nogueira2019passage,craswell2020overview, ren2021rocketqav2}.
Additionally, we evaluate likelihood-based retrieval methods which use pre-trained language models (PLMs) to estimate the probability of the query conditioned on variations of the corpus \citep{sachan2022improving}, such as measuring the query likelihood conditioned on the corpus with and without the candidate segment. 
Finally, we also evaluate the long context window \texttt{gpt-3.5-turbo-16k} ChatGPT model because of its ability to process long texts and perform instruction following.
We find that there is room for improvement on \tasknameLowercase{} across all methods.
For example, the bi-encoder systems  \citep{reimers2019sentence} struggle when the query is not semantically similar to the text segment that causes it; this often happens in the \conv{} and \lecture{} domain, where the query may be phrased differently than the original content.
Overall, our results indicate that \tasknameLowercase{} is a challenging task which requires new retrieval approaches to take in \textit{causal} relevance into account; 
for instance, the top-3 accuracy of the best model is only $44\%$ on the \lecture{} domain.

In summary, we make the following contributions in this paper: 

\begin{itemize}
    \item We propose a new task called \tasknameLowercase{} where the goal is to retrieve the cause of the query from a corpus. 
    This task targets the information need of \textit{content creators} who wish to improve their content in light of questions from \textit{information seekers}.
    \item We formalize a benchmark consisting of three domains for which \tasknameLowercase{} plays an important role in identifying the context triggering a user's query: retrieving the cause of student confusion in the \textsc{Lecture} setting, reader curiosity in the \textsc{News Article} setting, and user emotion in the \textsc{Conversation} setting.
    \item We evaluate a suite of popular retrieval systems, including bi-encoder and re-ranking architectures, as well as likelihood-based methods that use pretrained language models to estimate the probability of the query conditioned on variations of the corpus. 
    \item We show that there is room for improvement and limitations in current retrieval methods for performing \tasknameLowercase{}, suggesting that the task is not only challenging but also requires new retrieval approaches.
\end{itemize}

\begin{figure*}[ht]
    \centering 
   \includegraphics[width=\linewidth]{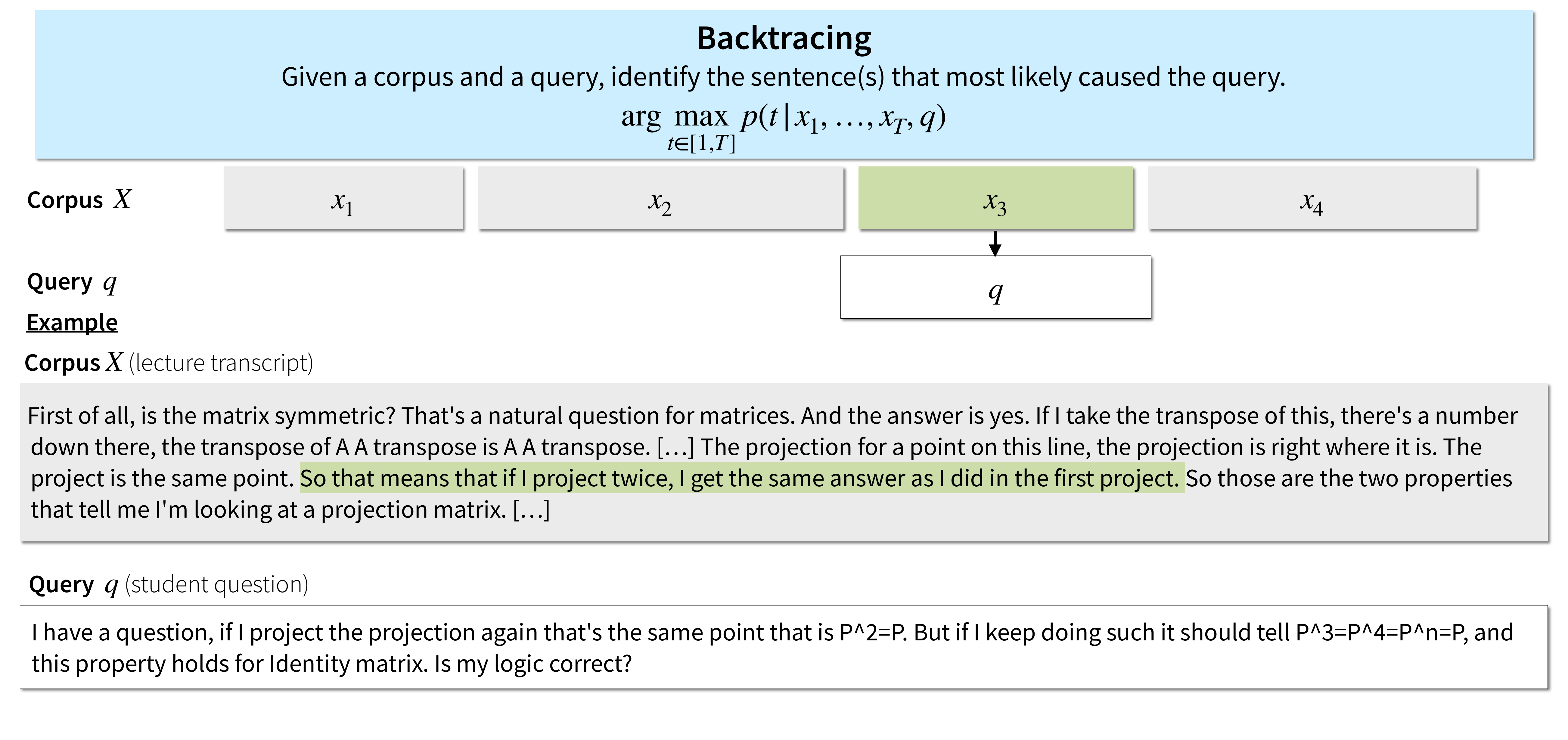}
   \vspace{-2em}
    \caption{
    \label{fig:backassociation}
    \small
    Illustration of \tasknameLowercase{}. 
    The goal of \tasknameLowercase{} is to identify the most likely sentence from the ordered corpus $X$ that caused the query $q$.  
    One example is the \lecture{} domain where the corpus is a lecture transcript and the query is a student question. 
    The lecturer only discusses about projecting twice and the student further extends that idea to something not raised in the lecture, namely into projecting a matrix an arbitrary $n$ times.
    } 
    \vspace{-0.5em}
\end{figure*}
\section{Related works}

The task of information retrieval (IR) aims to retrieve relevant documents or passages that satisfy the information need of a user \citep{schutze2008introduction, thakur2021beir}.
Prior IR techniques involve neural retrieval methods like ranking models \citep{guo2016deep,xiong2017end,khattab2020colbert} and representation-focused language models \citep{peters-etal-2018-deep, devlin2018bert, reimers2019sentence}.
Recent works also use PLMs for ranking texts in performing retrieval \citep{zhuang2021tilde, zhuang2021deep, sachan2022improving}; an advantage of using PLMs is not requiring any domain- or task-specific training, which is useful for settings where there is not enough data for training new models.
These approaches have made significant advancements in assisting \textit{information seekers} in accessing information on a range of tasks.
Examples of these tasks include recommending news articles to read for a user in the context of the current article they're reading \citep{voorhees2005trec, soboroff2018trec},  retrieving relevant bio-medical articles to satisfy health-related concerns \citep{tsatsaronis2015overview,boteva2016full, roberts2021overview,soboroff2021overview}, 
finding relevant academic articles to accelerate a researcher's literature search \citep{voorhees2021trec}, or extracting answers from texts to address questions \citep{yang2015wikiqa,rajpurkar2016squad,joshi2017triviaqa,yang2018hotpotqa}.

However, the converse needs of \textit{content creators} have received less exploration.
For instance, understanding what aspects of a lecture cause students to be confused remains under-explored and marks areas for improvement for content creators. 
\tasknameUppercase{} is related to work on predicting  search intents from previous user browsing behavior for understanding why users issue queries in the first place and what trigger their information needs \citep{cheng2010actively, kong2015predicting, koskela2018proactive}. 
The key difference between our approach and prior works is the nature of the input data and prediction task. 
While previous methods rely on observable user browsing patterns (e.g., visited URLs and click behaviors) for ranking future search results, our \tasknameLowercase{} framework leverages the language in the content itself as the context for the user query and the output space for prediction.
This shift in perspective allows content creators to get granular insights into specific contextual, linguistic triggers that influence user queries, as opposed to behavioral patterns.

Another related task is question generation, which also has applications to education \citep{heilman2010good, duan2017question, pan2019recent}.
While question generation settings assume the answer can be identified in the source document, \tasknameLowercase{} is interested in the triggers for the questions rather than the answers themselves.
In many cases, including our domains, the answer to the question may exist outside of the provided source document.

\section{\tasknameUppercase{} \label{sec:task_definition}}
Formally, we define \tasknameLowercase{} as:  Given corpus of $N$ sentences $X=\{x_1, \dots, x_N\}$ and query $q$, \tasknameLowercase{} selects
\begin{equation}
    \hat{t} = \arg \max_{t \in 1 \dots N} p(t|x_1, \dots, x_N,q)
\end{equation}
where $x_t$ is the $t^{th}$ sentence in corpus $X$ and $p$ is a probability distribution over the corpus indices, given the corpus and the query.
Figure~\ref{fig:backassociation} illustrates this definition and grounds it in our previous lecture domain example.
This task intuitively translates to: Given a lecture transcript and student question, retrieve the lecture sentence(s) that most likely caused the student to ask that question. 

Ideal methods for \tasknameLowercase{} are ones that can provide a continuous scoring metric over the corpus and can handle long texts. 
This allows for distinguishable contributions from multiple sentences in the corpus, as there can be more than one sentence that could cause the query. 
In the case where there is more than one target sentence, our acceptance criterion is whether there's overlap between the target sentences and the predicted sentence.
Additionally, some text domains such as lectures are longer than the context window lengths of existing language models. 
Effective methods must be able to circumvent this constraint algorithmically (e.g., by repeated invocation of a language model).

Our work explores the \tasknameLowercase{} task in a ``zero-shot'' manner across a variety of domains, similar to \citet{thakur2021beir}.
We focus on a restricted definition of zero-shot in which validation on a small development set is permitted, but not updating model weights. 
This mirrors many emerging real-world scenarios in which some data-driven interventions can be applied but not enough data is present for training new models. 
Completely blind zero-shot testing is notoriously hard to conduct within a reusable benchmark \citep{fuhr2018some,perez2021true} and is much less conducive to developing different methods, and thus lies outside our scope.

\section{\tasknameUppercase\label{sec:datasets} Benchmark Domains}
We use a diverse set of domains to establish a benchmark for  \tasknameLowercase{}, highlighting both its broad applicability and the shared challenges inherent to the task. 
This section first describes the domain datasets and then describes the dataset statistics with respect to the \tasknameLowercase{} task. 

\begin{table}[t]
  \centering
  \small
  \begin{tabular}{cc|c|c|c}
    \toprule
    \multicolumn{1}{c}{} & \multicolumn{1}{c}{} &  \multicolumn{1}{c}{\bf \textsc{Lec}} & \multicolumn{1}{c}{\bf \textsc{News} } &\multicolumn{1}{c}{\bf \textsc{Conv} }  \\ 
    % \multicolumn{1}{c}{ } & \multicolumn{1}{c}{} & \multicolumn{1}{c}{} & \multicolumn{1}{c}{} & \multicolumn{1}{c}{}  \\ 
    \midrule
    \multicolumn{1}{c}{Query} & \multicolumn{1}{c}{Total} & $210$ &  $1382$ & $671$ \\
    \multicolumn{1}{c}{} & \multicolumn{1}{c}{Avg. words} & $30.9$ &  $7.1$ & $11.6$ \\
    \multicolumn{1}{c}{} & \multicolumn{1}{c}{Max words} & $233$ &  $27$ & $62$ \\
    \multicolumn{1}{c}{} & \multicolumn{1}{c}{Min words} & $4$ &  $1$ & $1$ \\
    \midrule
    \multicolumn{1}{c}{Corpus} & \multicolumn{1}{c}{Total} & $11042$ &  $2125$ & $8263$ \\
    \multicolumn{1}{c}{} & \multicolumn{1}{c}{Avg. size} & $525.8$ &  $19.0$ & $12.3$ \\
    \multicolumn{1}{c}{} & \multicolumn{1}{c}{Max size} & $948$ &  $45$ & $6110$ \\
    \multicolumn{1}{c}{} & \multicolumn{1}{c}{Min size} & $273$ &  $ 7$ & $6$ \\
    \bottomrule
  \end{tabular}
  \caption{
  \small 
  Dataset statistics on the query and corpus sizes for \tasknameLowercase{}. 
  \textsc{Lec} is the \textsc{Lecture} domain, 
  \textsc{News} is the \textsc{News Article} domain, and 
  \textsc{Conv} is the \textsc{Conversation} domain. 
  The corpus size is measured on the level of sentences for \lecture{} and \news{}, and of conversation turns for \conv{}.
  }
  \label{tab:dataset_statistics}
\end{table}

% \begin{table}[t]
%   \centering
%   \small
%   \begin{tabular}{c|c|c|c}
%     \toprule
%     % \multicolumn{3}{c}{\bf }  &\multicolumn{9}{c}{\bf Accuracy} 
%     \multicolumn{1}{c}{\# sentences }   &\multicolumn{1}{c}{\bf Daily Dialog } &\multicolumn{1}{c}{\bf Inquisitive } &\multicolumn{1}{c}{\bf SIGHT} 
% \\ 
%     \midrule
%     \multicolumn{1}{c}{Mean}& $5.19$ & $7.83$ &  $18.8$ & $512$ \\
%     \multicolumn{1}{c}{Median}& $5$ &  $7.0$ &  $17$ & $484$\\
%     \multicolumn{1}{c}{Min}& $2$ & $5$ & $7$ &  $53$ \\
%     \multicolumn{1}{c}{Max}& $14$ & $18$ & $45$ & $1055$\\
%     \bottomrule
%   \end{tabular}
%   \caption{\small  Dataset statistics for backassociation. Number of distractor sentences.}
%   \label{tab:dataset_statistics}
% \end{table}

% inquisitive: 18.8 +/- 8.84
% Min: 7
% Max: 45
% Median: 17.0
% No. of queries: 1057

% squad: 5.19 +/- 2.06
% Min: 2
% Max: 14
% Median: 5.0
% sight: 512.03 +/- 157.46
% Min: 53
% Max: 1055
% Median: 484.0

\subsection{Domains \label{sec:domains}}
Figure~\ref{fig:backassociation_examples} illustrates examples of the corpus and query in each domain.
Table~\ref{tab:dataset_statistics} contains statistics on the dataset.
The datasets are protected under the CC-BY license.

\paragraph{\textsc{Lecture}}
We use real-world university lecture transcripts and student comments to construct the \textsc{Lecture} domain. 
Lectures are a natural setting for students to ask questions to express confusion about novel concepts. 
Lecturers can benefit from knowing what parts of their lecture cause confusion. 
We adapt the paired comment-lecture dataset from \textsc{Sight} \citep{wang2023sight}, which contains lecture transcripts from MIT OpenCourseWare math videos and real user comments from YouTube expressing confusion.
While these comments naturally act as queries in the \tasknameLowercase{} framework, the comments do not have ground-truth target annotations on what \textit{caused} the comment in the first place. 
Our work contributes these annotations. 
Two annotators (co-authors of this paper) familiar with the task of \tasknameLowercase{} and fluent in the math topics at a university-level annotate the queries\footnote{The annotators must be fluent in the math topics to understand both the lecture and query, and backtrace accordingly.}.
They select up to 5 sentences and are allowed to use the corresponding video to perform the task.
$20$ queries are annotated by both annotators and these annotations share high agreement:
the annotators identified the same target sentences for $70\%$ of the queries, and picked target sentences close to each other.
\textit{These annotation results indicate that performing \tasknameLowercase{} with consensus is possible.}
Appendix~\ref{app:sight_annotation_interface} includes more detail on the annotation interface and agreement.
The final dataset contains 210 annotated examples, comparable to other IR datasets \citep{craswell2020overview,craswell2021overview,soboroff2021overview}.\footnote{After conducting 2-means 2-sided equality power analysis, we additionally concluded that the dataset size is sufficiently large---the analysis indicated a need for 120 samples to establish statistically significant results, with power $1-\beta=0.8$ and $\alpha=0.05$.}
In the case where a query has more than one target sentence, the accuracy criterion is whether there's overlap between the target sentences and predicted sentence (see task definition in Section~\ref{sec:task_definition}).

\paragraph{\textsc{News Article}}
We use real-world news articles and questions written by crowdworkers as they read through the articles to construct the \textsc{News Article} domain. News articles are a natural setting for readers to ask curiosity questions, expressing a need for more information. 
We adapt the dataset from \citet{ko2020inquisitive} which contains news articles and questions indexed by the article sentences that provoked curiosity in the reader. 
We modify the dataset by filtering out articles that cannot fit within the smallest context window of models used in the likelihood-based retrieval methods (i.e., $1024$ tokens). 
This adapted dataset allows us to assess the ability of methods to incorporate more contextual information and handling more distractor sentences, while maintaining a manageable length of text. 
The final dataset contains 1382 examples.

\paragraph{\textsc{Conversation}} 
We use two-person conversations which have been annotated with emotions, such as \textit{anger} and \textit{fear}, and cause of emotion on the level of conversation turns.
Conversations are natural settings for human interaction where a speaker may accidentally say something that evokes strong emotions like anger.
These emotions may arise from cumulative or non-adjacent interactions, such as the example in Figure~\ref{fig:backassociation_examples}.
While identifying content that evokes the emotion expressed via a query differs from content that causes confusion, the ability to handle both is key to general and effective backtracing systems that retrieve information based on causal relevance.
Identifying utterances that elicit certain emotions can pave the way for better emotional intelligence in systems and refined conflict resolution tools. 
We adapt the conversation dataset from \citet{poria2021recognizing} which contain turn-level annotations for the emotion and its cause, and is designed for recognizing the cause of emotions.
The query is one of the speaker's conversation turn annotated with an emotion and the corpus is all of the conversation turns.
To ensure there are enough distractor sentences, we use conversations with at least 5 sentences and use the last annotated utterance in the conversation. 
The final dataset contains 671 examples.

\subsection{Domain Analysis \label{sec:domain_statistics}}

To contextualize the experimental findings in Section~\ref{sec:results}, we first analyze the structural attributes of our datasets in relation to \tasknameLowercase. 

\begin{figure}
    \centering
    \includegraphics[width=\linewidth]{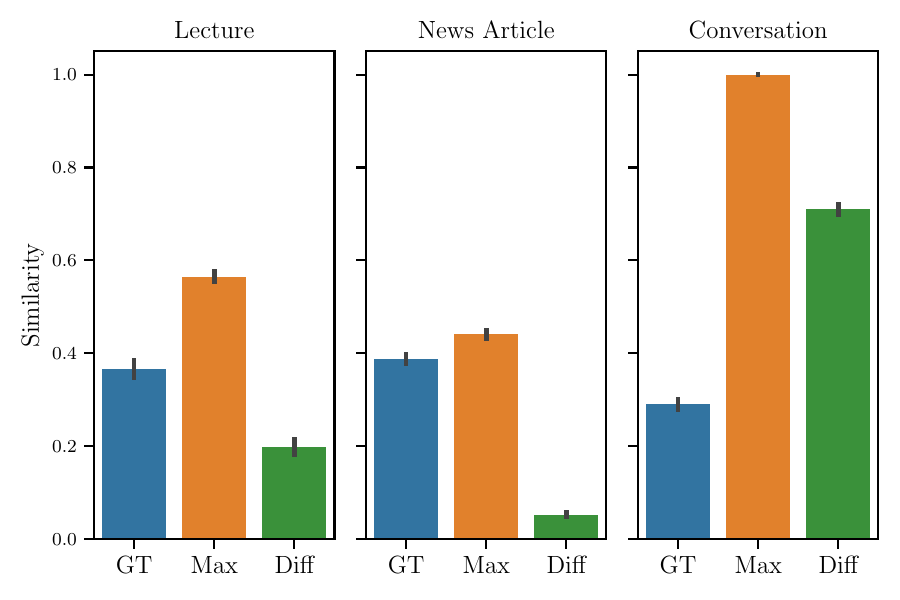}
    \caption{
    \small
    Each dataset plot shows the query similarity to the ground truth cause sentence (GT), to the corpus sentence with maximal similarity (Max), and the difference between the maximal and ground-truth similarity sentences (Diff).
    }
    \label{fig:similarity_comparison}
    \vspace{-1em}
\end{figure}

\paragraph{How similar is the query to the cause?}
To answer this question, we plot the semantic similarity of the query to the ground-truth cause sentence (GT) in Figure~\ref{fig:similarity_comparison}.
We additionally plot the maximal similarity of the query to any corpus sentence (Max) and the difference between the ground-truth and maximal similarity  (Diff). 
This compares the distractor sentences to the ground-truth sentences; the larger the difference is, the less likely semantic relevance can be used as a proxy for \textit{causal} relevance needed to perform \tasknameLowercase{}.
This would also indicate that poor performance of similarity-based methods because the distractor sentences exhibit higher similarity.
We use the \texttt{all-MiniLM-L12-v2} S-BERT model to measure semantic similarity \citep{reimers2019sentence}.

Notably, the queries and their ground-truth cause sentences exhibit low semantic similarity across domains, indicated by the low blue bars.
Additionally, indicated by the green bars, \conv{} and \lecture{} have the largest differences between the ground-truth and maximal similarity sentences, whereas \news{} has the smallest.
This suggests that there may be multiple passages in a given document that share a surface-level resemblance with the query, but a majority do not cause the query in the  \conv{} and \lecture{} domains.
In the \news{} domain, the query and cause sentence exhibit higher semantic similarity because the queries are typically short and mention the event or noun of interest. 
Altogether, this analysis brings forth a key insight: Semantic relevance  doesn't always equate causal relevance.

\paragraph{Where are the causes located in the corpus?}

\begin{figure}
    \centering
    \includegraphics[width=\linewidth]{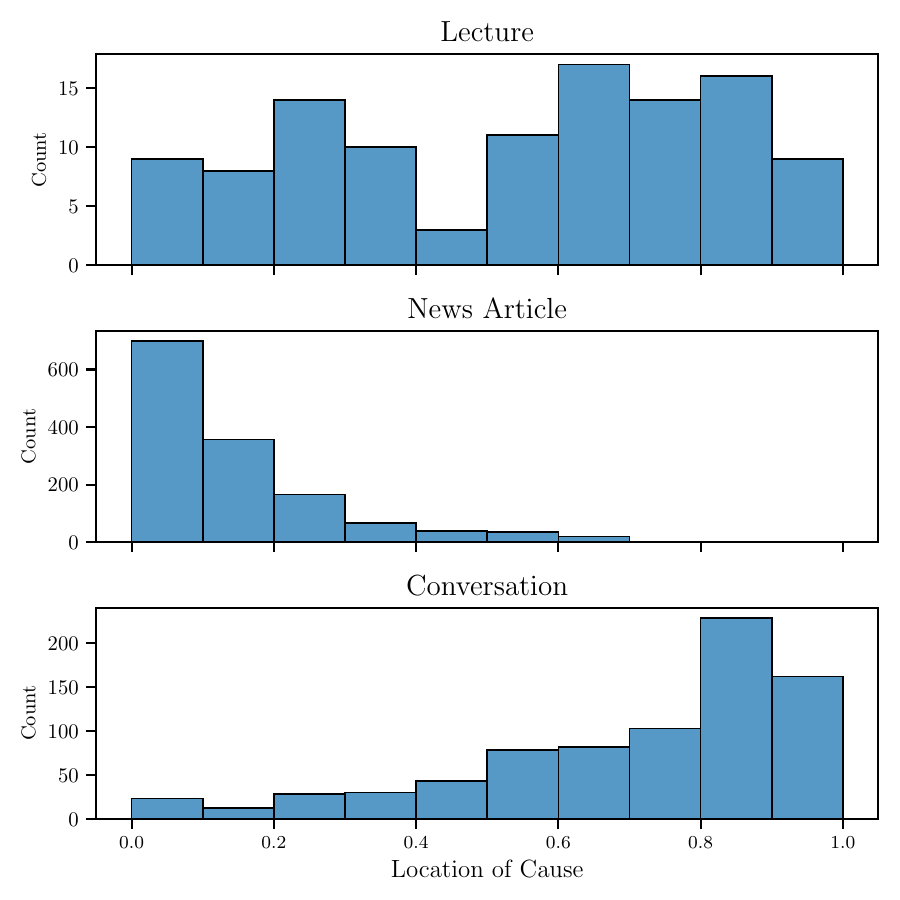}
    \caption{
    \small
    Each row plot is a per-domain histogram of where the ground-truth cause sentence lies in the corpus document.
    The x-axis reports the location of the cause sentence; $0$ means the cause sentence is the first sentence and $1$ the last sentence.
    The y-axis reports the count of cause sentences at that location.
    }
    \label{fig:cause_histogram}
    \vspace{-1em}
\end{figure}

Understanding the location of the cause  provides insight into how much context is needed in identifying the cause to the query. 
Figure~\ref{fig:cause_histogram} visualizes the distribution of cause sentence locations within the corpus documents.
These plots show that while some domains have causes concentrated in specific sections, others exhibit a more spread-out pattern. 
For the \news{} domain, there is a noticeable peak at the beginning of the  documents which suggests little  context is needed to identify the cause.
This aligns with the typical structure of news articles where crucial information is introduced early to capture the reader's interest. 
As a result, readers may have immediate questions from the onset.
Conversely, in the \conv{} domain, the distribution peaks at the end, suggesting that more context from the conversation is needed to identify the cause.
Finally, in the \lecture{} domain, the distribution is relatively uniform which suggests a broader contextual dependence. 
The causes of confusion arise from any section, emphasizing the importance of consistent clarity throughout an educational delivery.

An interesting qualitative observation is that there are shared cause locations for different queries. 
An example from the \lecture{} domain is shown in Figure~\ref{fig:common_confusion_point} where different student questions are  mapped to the same cause sentence. 
This shows the potential for models to effectively perform \tasknameLowercase{} and automatically identify common locations of confusion for lecturers to revise for future course offerings.

\begin{figure}
\fbox{\begin{minipage}{0.47\textwidth}
\small
\textbf{Lecture:} \textbf{[...]} So it's 1 by 2x0 times 2y0, which is 2x0y0, which is, lo and behold, 2. \textbf{[...]} \\ 
\textbf{Student A's question:} why is 2xo(yo) = 2?\\
\textbf{Student B's question:} When he solves for the area of the triangle, why does he say it doesn't matter what X0 and Y0 are? Does he just mean that all values of f(x) = 1/x will result in the area of the triangle of the tangent line to be 2? \\
\textbf{Student C's question:} Why always 2?? is there a prove?
\end{minipage}
}
% \vspace{-1em}
\caption{
% \small
An example of a common confusion point where several students posed questions concerning a particular part of the lecture.}
\label{fig:common_confusion_point}
\vspace{-2em}
\end{figure}

\section{Methods  \label{sec:methods}}
We evaluate a suite of existing, state-of-the-art retrieval methods and report their top-1 and top-3 accuracies: Do the top 1 and 3 candidate sentences include the ground-truth sentences? 
Reporting top-k accuracy is a standard metric in the retrieval setting.
We also report their minimum distance within the top-1 and top-3 candidates: What is the minimum distance between the method's candidates and the ground-truth sentences? 
The methods can be broadly categorized into similarity-based (i.e., using sentence similarity) and likelihood-based retrieval methods.
Similar to \citet{sachan2022improving}, the likelihood-based retrieval methods use PLMs to measure the probability of the query conditioned on variations of the corpus and can be more expressive than the similarity-based retrieval methods; we describe these variations in detail below.
We use GPT-2 \citep{radford2019language}, GPT-J \citep{wang2021gpt}, and OPT-6.7B \citep{zhang2022opt} as the PLMs. 
We additionally evaluate with \texttt{gpt-3.5-turbo-16k}, a new model that has a long context window ideal for long text settings like \textsc{Sight}. 
However, because this model does not output probability scores, we cast only report its top-1 results.

\paragraph{Random.}
This method randomly retrieves a sentence from the corpus.

\paragraph{Edit distance.}
This method retrieves the sentence with the smallest edit distance from the query.

\paragraph{Bi-encoders.}
This method retrieves the sentence with the highest semantic similarity using the best performing S-BERT models \citep{reimers2019sentencebert}. 
We use \texttt{multi-qa-MiniLM-L6-cos-v1} trained on a large set of question-answer pairs and \texttt{all-MiniLM-L12-v2} trained on a diversity of text pairs from sentence-transformers as the encoders.

\paragraph{Cross-encoder.}
This method picks the sentence with the highest predicted similarity score by the cross-encoder.
We use \texttt{ms-marco-MiniLM-L-6-v2}  \citep{thakur2021beir}.

\paragraph{Re-ranker.}
This method uses a bi-encoder to retrieve the top $k$ candidate sentences from the corpus, then uses a cross-encoder to re-rank the $k$ sentences.
We use \texttt{all-MiniLM-L12-v2} as the bi-encoder and \texttt{ms-marco-MiniLM-L-6-v2} as the cross-encoder. 
Since the smallest dataset---Daily Dialog---has a minimum of 5 sentences, we use $k=5$ for all datasets.

\paragraph{\texttt{gpt-3.5-turbo-16k}.}
This method is provided a line-numbered corpus and the query, and generates the line number that most likely caused the query. 
The prompt used for \texttt{gpt-3.5-turbo-16k} is in Appendix~\ref{app:scoring_prompts}.

\begin{spacing}{0.5}
\begin{table*}[t]
  \centering
  \small
  \resizebox{\textwidth}{!}{%
\def\arraystretch{.90}
  \begin{tabular}{cl|cc|cc|cc}
    \toprule
    \multicolumn{1}{c}{\bf}  &\multicolumn{1}{c}{\bf } &\multicolumn{2}{c}{\bf{\textsc{Lecture} }} &\multicolumn{2}{c}{\bf \textsc{News Article }}  &\multicolumn{2}{c}{\bf \textsc{Conversation}}
\\ 
    \multicolumn{1}{c}{} & \multicolumn{1}{c}{} &\multicolumn{1}{c}{ @1} &\multicolumn{1}{c}{ @3} &\multicolumn{1}{c}{ @1} &\multicolumn{1}{c}{ @3} &\multicolumn{1}{c}{ @1} &\multicolumn{1}{c}{ @3}\\
    \midrule
    & Random  & $0$ & $3$ &  $6$ & $21$  & $11$  & $31$  \\
    & Edit  & $4$ & $8$  & $7$ & $18$ & $1$ & $16$ \\
    & BM25  & $8$ & $15$  & $43$ & $65$ & $1$ & $35$ \\
    & Bi-Encoder (Q\&A)  & $23$ & $37$ & $48$ & $71$ & $1$ & $32$  \\
    & Bi-Encoder (\texttt{all-MiniLM}) & $26$ & $40$ & $49$ & $75$ & $1$  & $37$ \\
    & Cross-Encoder & $22$ & $39$ & $66$ & $\bf 85$ & $1$ &  $15$  \\
    & Re-ranker & $\bf 30$ &  $\bf 44$ & $66$ & $\bf 85$ & $1$  & $21$  \\
    & \texttt{gpt-3.5-turbo-16k} & $15$ & \text{N/A} & $\bf 67$ & \text{N/A}  & $ \bf 47$ & \text{N/A}  \\
    \midrule
    \bf Single-sentence & GPT2 & $21$ & $34$ & $43$ & $64$ & $3$ & $46$ \\
    $p(q|s_t)$ & GPTJ  & $23$ & $42$ & $\bf 67$ & $\bf 85$  & $5$ &  $ \bf  65$ \\
    & OPT 6B  & $\bf 30$ & $ 43$ & $66$ & $82$  & $2$ & $56$ \\
    \midrule
    \bf Autoregressive & GPT2  & $11$ & $16$ & $9$ & $18$  & $5$ & $54$\\
    $p(q|s_{\leq t})$ & GPTJ  & $14$ & $24$  &  $55$ &  $76$ & $8$ & $60$ \\
    & OPT 6B  & $16 $ & $26$ &  $52$ &  $73$   & $18$ & $ \bf  65$ \\
    \midrule
    \bf ATE  & GPT2  & $13 $ & $21$ &  $51$ &  $68$& $2$ & $24$ \\
    $p(q|S) - p(q|S /\ \{s_t\}\ )$ & GPTJ  &  $8$ & $18$ &  $\bf 67$ &  $79$& $3$ & $18$ \\
    & OPT 6B  & $2$ & $6$ & $64$ & $76$  & $3$ & $22$ \\
    \bottomrule
  \end{tabular}
  }
  \caption{
  % \small 
    \textbf{Accuracy ($\uparrow$ \% betterd).} The best models in each column are bolded. 
    For each dataset, we report the top-$1$ and $3$ accuracies.
    \gptLong{} reports N/A  for top-3 accuracy because it does not output deterministic continuous scores for ranking sentences.
    }
  \label{tab:accuracy_evaluation}
    \vspace{-0.75em}
\end{table*}
\end{spacing}

\begin{spacing}{0.5}
\begin{table*}[t]
  \centering
  \small
  \resizebox{\textwidth}{!}{%
\def\arraystretch{.90}
  \begin{tabular}{cl|cc|cc|cc}
    \toprule
    \multicolumn{1}{c}{\bf}  &\multicolumn{1}{c}{\bf } &\multicolumn{2}{c}{\bf{\textsc{Lecture} }} &\multicolumn{2}{c}{\bf \textsc{News Article }}  &\multicolumn{2}{c}{\bf \textsc{Conversation}} \\
    \multicolumn{2}{c}{\bf } &\multicolumn{1}{c}{ @1} &\multicolumn{1}{c}{ @3} &\multicolumn{1}{c}{ @1} &\multicolumn{1}{c}{ @3} &\multicolumn{1}{c}{ @1} &\multicolumn{1}{c}{ @3}\\
    \midrule
    & Random  & $167.5$ & $67.8 $ &  $7.6 $ & $3.0 $  & $3.7$  & $1.7$  \\
    & Edit  & $157.9 $ & $70.7 $ &  $7.7 $ & $3.4 $  & $1.3$  & $0.9$  \\
    & BM25  & $122.7 $ & $50.7 $ &  $4.6 $ & $1.4 $  & $1.3$  & $0.7$  \\
    & Bi-Encoder (Q\&A)  & $ 91.9$ & $35.2 $ &  $ 4.1$ & $ 1.2$  & $1.3$  & $0.8$  \\
    & Bi-Encoder (\texttt{all-MiniLM}) & $ 84.7$ & $ 38.6$ &  $3.7 $ & $1.0 $  & $1.3$  & $ 0.7$  \\
    & Cross-Encoder & $96.6 $ & $33.8 $ &  $2.5 $ & $ \bf 0.6 $  & $1.3 $  & $ 0.9$  \\
    & Re-ranker & $92.2$ & $41.4 $ &  $2.7 $ & $\bf 0.6 $  & $ 1.3$  & $0.9$  \\
    & \texttt{gpt-3.5-turbo-16k} & $  73.9$ & N/A &  $\bf 1.5 $ & N/A  & $\bf  1.0$  & N/A  \\
    \midrule
    \bf Single-sentence & GPT2 & $ 5.4^*$ & $2.1^* $ &  $ 4.6$ & $1.5 $  & $1.5$  & $0.6$  \\
    $p(q|s_t)$ & GPTJ  & $\bf 5.0^*$ & $\bf  1.9^*$ &  $ 2.5$ & $ 0.7$  & $ 1.4$  & $\bf  0.4$  \\
    & OPT 6B  & $ 5.2^*$ & $ 2.3^*$ &  $ 2.7$ & $0.8 $  & $ 1.3$  & $0.5 $  \\
    \midrule
    \bf Autoregressive & GPT2  & $5.6^* $ & $3.4^*$ &  $ 7.2$ & $ 4.8$  & $2.0 $  & $ 0.8$  \\ 
    $p(q|s_{\leq t})$ & GPTJ  & $5.5^*$ & $3.4^*$ &  $1.8 $ & $ 0.8$  & $ 2.0$  & $ 0.8$  \\
    & OPT 6B  & $ 5.1^* $ & $3.5^* $ &  $1.9 $ & $ 1.0$  & $ 1.9$  & $ 0.7$  \\
    \midrule
    \bf ATE  & GPT2  & $7.4^* $ & $2.8^* $ &  $  4.7$ & $1.3 $  & $ 1.5 $  & $ 0.9$  \\
    $p(q|S) - p(q|S /\ \{s_t\}\ )$ & GPTJ  & $ 7.2^*$ & $3.2^* $ &  $ 2.9 $ & $0.9 $  & $ 1.6 $  & $ 1.0 $  \\
    & OPT 6B  & $7.1^* $ & $ \bf  1.9^* $ &  $3.2 $ & $  1.1$  & $2.4 $  & $1.0 $  \\
    \bottomrule
  \end{tabular}
  }
  \caption{
  % \small 
  \textbf{Minimum Sentence Distance from Ground Truth ($\downarrow$ better)}
    The best models in each column are bolded. 
    For each dataset, we report the minimum sentence distance from the ground truth cause sentence of the method's top-$1$ and $3$ candidates; 0 meaning that the method always predicts the ground truth candidate sentence. 
    Note for the likelihood-based methods on the \textsc{Lecture} domain were evaluated on 20-sentence chunks of the original text due to the context window limitation. 
    If the top sentence is not in the top-chunk, it is excluded in distance metric.  
    We've marked the affected metrics with an asterisk $^*$.
    }
  \label{tab:distance_evaluation}
    \vspace{-0.75em}
\end{table*}
\end{spacing}

\paragraph{Single-sentence likelihood-based retrieval $p(q|x_t)$.}
This method retrieves the sentence $x_t \in X$ that maximizes $p(q|x_t)$.
To contextualize the corpus and query, we add domain-specific prefixes to the corpus and query. 
For example, in \textsc{Sight}, we prepend ``Teacher says: '' to the corpus sentence and ``Student asks: '' to the query. 
Due to space constraints, Appendix~\ref{app:scoring_prompts} contains all the prefixes used.

\paragraph{Auto-regressive likelihood-based retrieval $p(q|x_{\leq t})$.}
This method retrieves the sentence $x_t$ which maximizes $p(q|x_{\leq t})$.
This method evaluates the importance of preceding context in performing \tasknameLowercase{}. 
\lecture{} is the only domain where the entire corpus cannot fit into the context window. 
This means that we cannot always evaluate $p(q|x_{\leq t})$ for $x_t$ when $|x_{\leq t}|$ is longer than the context window limit.
For this reason, we split the corpus $X$ into chunks of $k$ sentences, (i.e., $X_{0:k-1}, X_{k:2k-1}, \dots$) and evaluate each $x_t$ within their respective chunk. 
For example, if $x_t \in X_{k:2k-1}$, the auto-regressive likelihood score for $x_t$ is $p(q|X_{k:t})$.
We evaluate with $k=20$ because it is the maximum number of sentences (in addition to the query) that can fit in the smallest model context window.

\paragraph{Average Treatment Effect (\causalAcronym{}) likelihood-based retrieval $p(q|X)-p(q|X \setminus x_{t})$.}
This method takes inspiration from treatment effects in causal inference \citep{holland1986statistics}. 
We describe how \causalAcronym{} can be used as a retrieval criterion. 
In our setting, the treatment is whether the sentence $x_t$ is included in the corpus. 
We're interested in the effect the treatment has on the query likelihood: 
\begin{align}
    \texttt{\causalAcronym{}}(x_t) &= p_{\theta}(q|X) - p_{\theta}(q|X \setminus \{x_t\}).
    \label{eq:ate}
\end{align}

\causalAcronym{} likelihood methods retrieve the sentence that maximizes $\texttt{\causalAcronym{}}(x_t)$.
These are the sentences that have the largest effect on the query's likelihood. 
We directly select the sentences that maximize Equation~\ref{eq:ate} for \news{} and \conv{}.
We perform the same text chunking for \lecture{} as in the auto-regressive retrieval method: If $x_t \in X_{k:2k-1}$, the \causalAcronym{} likelihood score for $x_t$ is measured as $p(q|X_{k:2k-1}) - p(q|X_{k:2k-1} \setminus \{x_t\})$.

\section{Results \label{sec:results}}

The accuracy results are summarized in Table~\ref{tab:accuracy_evaluation}, and distance results in Table~\ref{tab:distance_evaluation}.

\paragraph{The best-performing models achieve modest accuracies.}
For example, on the \lecture{} domain with many distractor sentences, the best-performing model only achieves top-3 $44\%$ accuracy. 
On the \conv{} domain with few distractor sentences, the best-performing model only achieves top-3 $65\%$ accuracy.
This underscores that measuring causal relevance is challenging and markedly different from existing retrieval tasks. 

\paragraph{No model performs consistently across domains.}
For instance, while a similarity-based method like the Bi-Encoder (\texttt{all-MiniLM}) performs well on the \news{} domain with top-3 $75\%$ accuracy, it only manages top-3 $37\%$ accuracy on the \conv{} domain.
These results complement the takeaway from the domain analysis in Section~\ref{sec:datasets} that semantic relevance is not a reliable proxy for  causal relevance. % between a query and its triggering context.
Interestingly, on the long document domain \lecture{}, the long-context model \gptLong{} performs worse than non-contextual methods like single-sentence likelihood methods.
This suggests that accounting for context is challenging for current models. 

\paragraph{Single-sentence methods generally outperform their autoregressive counterparts except on \conv{}.}
This result complements the observations made in  Section~\ref{sec:datasets}'s domain analysis where the location of the causes concentrates at the start for \news{} and uniformly for \lecture{}, suggesting that little context is needed to identify the cause. 
Conversely, conversations require more context to distinguish the triggering contexts, which suggests why the autoregressive methods perform generally better than the single-sentence methods.

\paragraph{ATE likelihood methods does not signicantly improve upon other methods.}
Even though the ATE likelihood method is designed the calculate the effect of the cause sentence, it competes with noncontextual methods such as the single-sentence likelihood methods.
This suggest challenges in using likelihood methods to measure the counterfactual effect of a sentence on a query.

\section{Conclusion}
In this paper, we introduce the novel task of \tasknameLowercase{}, which aims to retrieve the text segment that most likely provokes a query.
This task addresses the information need of \textit{content creators} who want to improve their content, in light of queries from information seekers. 
We introduce a benchmark that covers a variety of domains, such as the news article and lecture setting.
We evaluate a series of methods including popular IR methods, likelihood-based retrieval methods and \gptLong{}.
Our results indicate that there is room for improvement across existing retrieval methods.
These results suggest that \tasknameLowercase{} is a challenging task that requires new retrieval approaches with better contextual understanding and reasoning about causal relevance.
We hope our benchmark serves as a foundation for improving future retrieval systems for \tasknameLowercase{}, and ultimately, spawns systems that empower content creators to understand user queries, refine their content and provide users with better experiences.

\section*{Limitations}

\paragraph{Single-sentence focus.} 
Our approach primarily focuses on identifying the most likely single sentence that caused a given query. 
However, in certain scenarios, the query might depend on groups or combinations of sentences.
Ignoring such dependencies can limit the accuracy of the methods.

\paragraph{Content creators in other domains.} Our evaluation primarily focuses on the dialog, new article and lecture settings.
While these domains offer valuable insights, the performance of \tasknameLowercase{} methods may vary in other contexts, such as scientific articles and queries from reviewers.
Future work should explore the generalizability of \tasknameLowercase{} methods across a broader range of domains and data sources.

\paragraph{Long text settings.} 
Due to the length of the lecture transcripts, the transcripts had to be divided and passed into the likelihood-based retrieval methods.
This approach may result in the omission of crucial context present in the full transcript, potentially affecting the accuracy of the  likelihood-based retrieval methods.
Exploring techniques to effectively handle larger texts and overcome model capacity constraints would be beneficial for improving backtracing performance in long text settings, where we would imagine \tasknameLowercase{} to be useful in providing feedback for.

\paragraph{Multimodal sources.} 
Our approach identifies the most likely text segment in a corpus that caused a given query. 
However, in multimodal settings, a query may also be caused by other data types, e.g., visual cues that are not captured in the transcripts.
Ignoring such non-textual data can limit the accuracy of the methods. 

\section*{Ethics Statement}
Empowering content creators to refine their content based on user feedback contributes to the production of more informative materials.
Therefore, our research has the potential to enhance the educational experiences of a user, by assisting content creators through \tasknameLowercase{}.
Nonetheless, we are mindful of potential biases or unintended consequences that may arise through our work and future work. 
For example, the current benchmark analyzes the accuracy of \tasknameLowercase{} on English datasets and uses PLMs trained predominantly on English texts. 
As a result, the inferences drawn from the current \tasknameLowercase{} results or benchmark may not accurately capture the causes of multilingual queries, and should be interpreted with caution.
Another example is that finding the cause for a user emotion can be exploited by content creators.
We consider this as an unacceptable use case of our work, in addition to attempting to identify users in the dataset or using the data for commercial gain.

\section*{Acknowledgements}
We'd like thank Michael Zhang and Dilip Arumugam for the fruitful conversations at the start of this project.
We'd also like to thank Gabriel Poesia for helpful feedback on the paper.

\bibliography{anthology, custom}

\appendix

\section{Computational Setup}
We ran our experiments on a Slurm-based university compute cluster on A6000 machines.
The experiments varied in length in time---some took less than an hour to run (e.g., the random baselines), while others took a few days to run (e.g., the ATE likelihood-based methods on \lecture{}).

\section{\textsc{Lecture} annotation interface \label{app:sight_annotation_interface}}

\begin{figure*}
    \centering 
    \includegraphics[width=\linewidth]{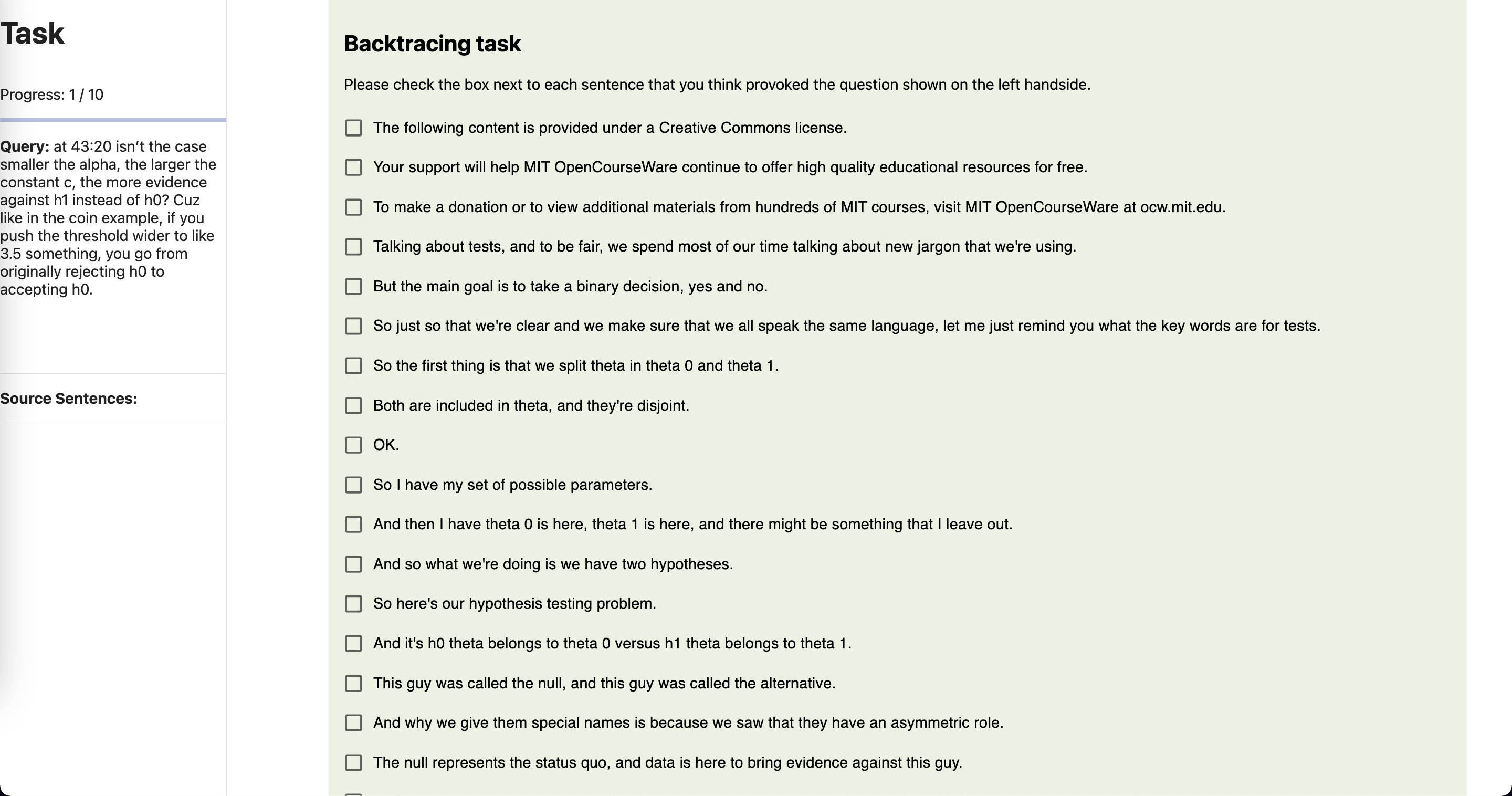}
    \caption{Annotation interface \label{fig:annotation_interface}}
\end{figure*}

Figure~\ref{fig:annotation_interface} shows the interface used for annotating the \textsc{Lecture} dataset. 

\section{Contextualized prefixes for scoring}
\label{app:scoring_prompts}
This section describes the prompts used for the likelihood-based retrieval methods and \gptLong{}. 

The prompts used for \gptLong{} follow the practices in works from NLP, education and social sciences \citep{inverse2022, minichain2023, ziems2023can, wang2023sight}. 
Specifically, we enumerate the sentences in the corpus as multiple-choice options and each option is separated by a newline.
We add context for the task at the start of the prompt, and the constraints of outputting a JSON-formatted text for the task at the end of the prompt.
We found the model to be reliable in outputting the text in the desirable format.

\subsection{\textsc{Lecture}}
For the likelihood-based retrieval methods, the sentences are concatenated by spaces and ``A teacher is teaching a class, and a student asks a question.$\backslash$nTeacher:  '' is prepended to the corpus. 
Because the text comes from transcribed audio which is not used in training dataset of the PLMs we use in our work, we found it important for additional context to be added in order for the probabilities to be slightly better calibrated. 
For the query, ``Student: '' is prepended to the text.
For example, $X=$ ``A teacher is teaching a class, and a student asks a question.$\backslash$n Teacher: [sentence 1] [sentence 2] ...'', and $q=$ ``Student: [query]''.

The prompt used for \gptLong{} is in Figure~\ref{fig:sight_prompt}.

\def\gptLong{\texttt{\texttt{gpt-3.5-turbo-16k}}}

\begin{figure*}[t]
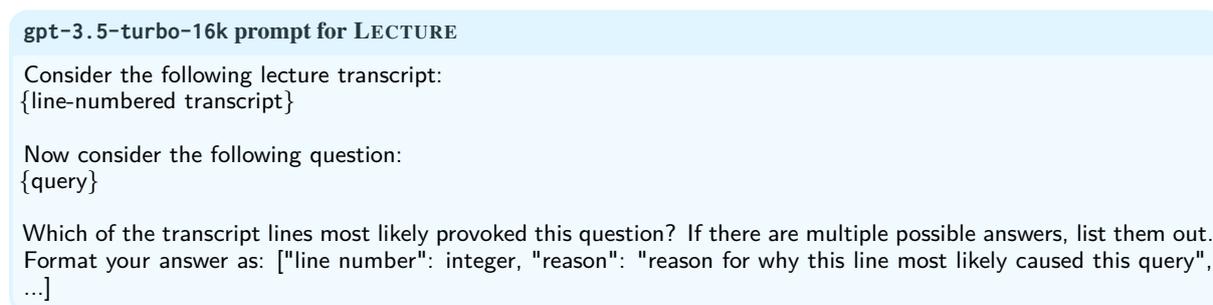

    \centering % Center the textbox within the figure environment
    \begin{tcolorbox}[
    sightsetting,
    title={\small \textbf{\gptLong{} prompt for \textsc{Lecture}}}
    ]
    \small 
    Consider the following lecture transcript: \\
    \{line-numbered transcript\} \\
    
    Now consider the following question: \\
    \{query\} \\
    
    Which of the transcript lines most likely provoked this question? If there are multiple possible answers, list them out. Format your answer as: [{{"line number": integer, "reason": "reason for why this line most likely caused this query"}}, ...]

    \end{tcolorbox}
    \vspace{-1em}
    \caption{
    \gptLong{} prompt for \textsc{Lecture}. For the line-numbered transcript, ``Teacher: '' is prepended to each sentence, the sentences are separated by line breaks, and each line begins with its line number. For the query, ``Student: '' is prepended to the text. For example, a line-numbered article looks like ``0. Teacher: [sentence 1]$\backslash$n1. Teacher: [sentence 2]$\backslash$n2. Teacher: [sentence 3] ...'', and the query looks like ``Student: [query]''.
    \label{fig:sight_prompt}
    }
\end{figure*}
\def\gptLong{\texttt{\texttt{gpt-3.5-turbo-16k}}}

\begin{figure*}[t]
    \centering % Center the textbox within the figure environment
    \begin{tcolorbox}[
    inquisitivesetting,
    title={\small \textbf{\gptLong{} prompt for \textsc{News Article}}}
    ]
    \small 
    Consider the following article: \\
    \{line-numbered article\} \\
    
    Now consider the following question: \\
    \{query\} \\
    
    Which of the article lines most likely provoked this question? If there are multiple possible answers, list them out. Format your answer as: [{{"line number": integer, "reason": "reason for why this line most likely caused this query"}}, ...]

    \end{tcolorbox}
    \vspace{-1em}
    \caption{
    \gptLong{} prompt for \textsc{News Article}. For the line-numbered article, ``Text: '' is prepended to each sentence, the sentences are separated by line breaks, and each line begins with its line number. For the query, ``Question: '' is prepended to the text. For example, a line-numbered article looks like ``0. Text: [sentence 1]$\backslash$n1. Text: [sentence 2]$\backslash$n2. Text: [sentence 3] ...'', and the query looks like ``Question: [question]''.
    \label{fig:inquisitive_prompt}
    }
\end{figure*}
\def\gptLong{\texttt{\texttt{gpt-3.5-turbo-16k}}}

\begin{figure*}[t]
    \centering % Center the textbox within the figure environment
    \begin{tcolorbox}[
    dailydialogsetting,
    title={\small \textbf{\gptLong{} prompt for \textsc{Conversation}}}
    ]
    \small 
    Consider the following conversation: \\
    \{line-numbered conversation\} \\
    
    Now consider the following line: \\
    \{query\} \\
    
    The speaker felt \{emotion\} in this line.
    Which of the conversation turns (lines) most likely caused this emotion? If there are multiple possible answers, list them out. Format your answer as: [{{"line number": integer, "reason": "reason for why this line most likely caused this emotion"}}, ...] 

    \end{tcolorbox}
    \vspace{-1em}
    \caption{
    \gptLong{} prompt for \textsc{Conversation}. For the line-numbered conversation, the speaker is added to each turn, the turns are separated by line breaks, and each line begins with its line number. For the query, the speaker is also added. For example, a line-numbered conversation may look like ``0. Speaker A: [utterance]$\backslash$n1. Speaker B: [utterance]$\backslash$n2. Speaker A: [utterance] ...'', and the query may look like ``Speaker A: [query]''.
    \label{fig:daily_dialog_prompt}
    }
\end{figure*}

\subsection{\textsc{News Article}}
For the likelihood-based retrieval methods, the sentences are concatenated by spaces and ``Text: '' is prepended to the corpus.
For the query, ``Question: '' is prepended to the text.
For example, $X=$ ``Text: [sentence 1] [sentence 2] ...'', and $q=$ ``Question: [question]''.

The prompt used for \gptLong{} is in Figure~\ref{fig:inquisitive_prompt}.

\subsection{\textsc{Conversation}}
For the likelihood-based retrieval methods, the speaker identity is added to the text, and the turns are separated by line breaks. 
For the query, the same format is used.
For example, $X=$ ``Speaker A: [utterance]$\backslash$nSpeaker B: [utterance]'', and $q=$ ``Speaker A: [query]''.

The prompt used for \gptLong{} is in Figure~\ref{fig:daily_dialog_prompt}. 

\end{document}